\documentclass[letter, 11pt]{article}
\usepackage[latin1]{inputenc}
\usepackage{amstext}
\usepackage{graphicx}
\usepackage{amsfonts}
\usepackage{latexsym}
\usepackage{amssymb}
\usepackage{amsthm}
\usepackage{amsmath}
\usepackage[margin=3.8cm,bottom=3.6cm,top=3.6cm]{geometry}

\title{The Evolution of Navigable Small-World Networks}
\author{Oskar Sandberg \thanks{The Department of Mathematical Sciences, Chalmers
    Technical University and Gothenburg University. ossa@math.chalmers.se},
Ian Clarke \thanks{FreenetProject Inc. ian@freenetproject.org}}

\date{\today}

\begin{document}

\newtheorem{theorem}{Theorem}[section]
\newtheorem{lemma}[theorem]{Lemma}
\newtheorem{definition}[theorem]{Definition}
\newtheorem{proposition}[theorem]{Proposition}
\newtheorem{corollary}[theorem]{Corollary}
\newtheorem{conjecture}[theorem]{Conjecture}
\newtheorem{algorithm}[theorem]{Algorithm}

\newcommand{\fs}{\mbox{ for some }}
\newcommand{\PR}{\mathbf{P}}
\newcommand{\given}{\,\mathbf{|}\,}

\newcommand{\E}{\mathbf{E}}
\newcommand{\R}{\mathbb{R}}
\newcommand{\Z}{\mathbb{Z}}

\maketitle

\begin{abstract}
  Small-world networks, which combine randomized and structured
  elements, are seen as prevalent in nature. Several random graph
  models have been given for small-world networks, with one of the
  most fruitful, introduced by Jon Kleinberg
  \cite{kleinberg:smallworld}, showing in which type of graphs it is
  possible to route, or navigate, between vertices with very little
  knowledge of the graph itself.

  Kleinberg's model is static, with random edges added to a fixed
  grid. In this paper we introduce, analyze and test a randomized
  algorithm which successively rewires a graph with every application.
  The resulting process gives a model for the evolution of small-world
  networks with properties similar to those studied by Kleinberg.
\end{abstract}

\thispagestyle{empty}

\newpage

\setcounter{page}{1}

\section{Introduction}

The ``Small World Phenonomenon'' is the name given to the observation
that seemingly random people can often find a short chain of
aquaintances connecting them to one another.  Mathematically, this has
been related \cite{watts:smallworld} \cite{newman:renorm} to the
observation that structured graphs, such as grids, can have their
diameter drastically reduced by the introduction on some random edges
between the vertices (as proved for the circle in
\cite{bollobas:diameter}).


Connected with this is the question, raised by Jon Kleinberg in 2000
\cite{kleinberg:smallworld}, whether short paths can be found between
any two vertices by actors in the network lacking global information
about the graph to use when routing. He showed that this is not
possible in all families of random graphs with small diameter, but
instead depends on very specific properties of certain classes of such
graphs.  Graphs where short paths can be found are often called
``navigable''.

The question of whether graphs are navigable is of particular
practical interest due to a multitude of applications. Specifically,
the type of routing suggested by Kleinberg has been employed in 
distributed computing, hashtables \cite{manku:symphony} and
peer-to-peer software \cite{clarke:freenet}.

\subsection{Motivation}

While previous results go a long way towards characterizing when
graphs are navigable, they leave open the question of how such graphs
are formed. At the same time, experiments with social networks (e.g.
\cite{milgram:smallworld} \cite{dodds:social}), seem to
indicate that those do, to at least some extent, have navigable
properties. A model for evolution and growth of navigable graphs,
similar in some respects to the preferential-attachment models of
power-law degree distributions \cite{bollobas:scalefree}
\cite{barabasi:emergence}, would help explain when and how they arise
through natural processes. Such a model could also be used to generate
graphs for use in networks where efficient routing is important, such
as the types of overlay networks on the Internet mentioned above. 

In a recent summary paper \cite{kleinberg:networks} Kleinberg
identifies this problem as one of central open issues in the area.

\subsection{Contribution}

We summarize our contributions as follows:
\begin{enumerate}
\item We present an evolving random graph model where the edges of a
  graph are re-wrired by performing repeated greedy walks between random
  points, and changing the edges based on these.
\item We analyze rigourously, under a few simplifications, the
  stationary case of the model, showing that it is a navigable random
  graph.
\item We simulate the algorithm in a number of different
  circumstances, showing that it leads to graphs that perform as well
  or better then those produced with Kleinberg's model.
\end{enumerate}

\subsection{Previous Work}

In a followup to his original work \cite{kleinberg:dynamics}, Jon
Kleinberg motivated why the necessary distribution for navigability
might arise in nature by means of ``group memberships''. He showed
that in a more generalized setting, structures are navigable if two
vertices are connected with a probability that is inversely
proportional the size of the smallest group they both populate. That
this should be the case is in some sense natural, since the
probability of knowing somebody may decrease with the size of the
group in which you know them. Similar arguments can be found in
\cite{nowell:geographic} and \cite{watts:social}.

A preprint paper by Clauset and Moore \cite{clauset:navig} presents a
different re-wiring algorithm for the creation of navigable graphs.
They show positive results for this algorithm using simulation, but do
not present any analytic results. In \cite{eppstein:steady} a
re-wiring algorithm for the creation of so called scale-free (or
power-law) graphs is presented. This does not deal with clustering nor
navigability, and no analytic results regarding the stationary
distribution are derived.

Early versions of the Freenet peer-to-peer data network, presented in
\cite{clarke:protecting} and \cite{clarke:freenet}, used a method
similar to the algorithm we propose to update the links between peers.
The current work is in part inspired by trying to apply the ideas from
the design of Freenet to an environment more conductive to analysis.
\cite{zhang:freenet} previously related Freenet to the discussion of
navigable small-world graphs, but they worked mostly on proposing
modifications to the algorithm that resulted in a more robust network,
instead of looking more closely at the properties of Freenet's
neighbor sampling.

\section{Navigable Small Worlds}

In his initial study of navigable graphs \cite{kleinberg:smallworld},
Kleinberg studied graphs constructed by starting with a two
dimensional grid, and adding random long-range contacts according to a
certain class of distributions. For the purpose of vertex to vertex
routing in such graphs, he defined a decentralized algorithm as one
where each vertex has to make a routing decision based only on the
grid positions of the query's destination and the vertex's immediate
neighborhood\footnotemark{}.

\footnotetext{For the upper bounds, he also allowed vertices to know the
  grid position of all previous vertices in the query and their
  neighbors. This was not used in the lower bounds, and so strengthens
both results.}

Kleinberg showed that if one starts with a two dimensional grid and
adds long-range edges between vertices $x$ and $y$ with probability
$\propto |x - y|^{\alpha}$ where $|x - y|$ denotes Manhattan-distance
in the grid, then only the case $\alpha = -2$ allows for decentralized
routing in a polylogarithmic number of steps. For all other values of
$\alpha$ a lower bound which is a fractional power of the graph size
can be derived.

In the critical case $\alpha = -2$, however, it is sufficient to use
the most direct routing possible, so called \emph{greedy routing}. As
the name implies, greedy routing means that at each step, one attempts to
minimize the distance to the destination. That is, if $x$ wishes to
route a query to vertex $z$, then he picks as the next step the one of
his neighbors (long-range or otherwise) which is closest to $z$. If
$n$ is the size of the graph, then a bound of $O(\log^2 n)$ on the
expected number of steps needed can be found. Kleinberg's model can
easily be extended to graphs formed by adding long-range edges to
grids of dimension other than two. If the basic grid has dimension
$d$, it can be seen that $\alpha = -d$ corresponds to the critical
case in which routing is possible.

\section{The Algorithm}

We let $V$ be the set of vertices, each with a position in a grid or
some other regular lattice. We will let $E$ be set of directed
shortcut (long-range) edges between vertices in $V$, and $G = (V,E)$
the resulting digraph. Let $G'$ be $G$ augmented by additional edges
going both ways between each pair of vertices that are adjacent in the
lattice. The proposed algorithm, which we call \emph{destination
  sampling}, is as follows:

\begin{algorithm}  
  Let $G_s = (V, E_s)$ be the directed graph of shortcuts at time $s$.
  Let $0 < p < 1$. Then $G_{s+1}$ is defined as follows.

\begin{enumerate}
\item Choose $y_{s+1}$ and $z_{s+1}$ uniformly from $V$.
\item If $y_{s+1} \neq z_{s+1}$, do a greedy walk in $G_s'$ from $y_s$
  to $z_s$ along the lattice and the shortcuts of $E_s$. Let $x_0 =
  y_{s+1}, x_1, x_2, ..., x_t = z_{s+1}$ denote the points of this
  walk.
\item For each $x_0, x_1,...,x_{t-1}$ with at least one shortcut,
  independently with probability $p$ replace a randomly chosen
  shortcut with one to $z_{s+1}$.
\end{enumerate}
\label{alg:main}
\end{algorithm}

After a walk is made, $G_{s+1}$ is the same as $G_s$, except that a
shortcut from each vertex in walk $s+1$ is with probability $p$
replaced by an edge to the destination. In this way, the destination
of each edge is a sample of the destinations of previous walks passing
through it. The claim is that updating the shortcuts using this
algorithm eventually results in a shortcut graph with greedy
path-lengths of $O(\log^2 n)$.

The value of $p$ is a parameter in the algorithm. It serves to
disassociate the shortcut from a vertex with that of its neighbors.
For this purpose, the lower the value of $p > 0$ the better, but very
small values of $p$ will also lead to slower sampling. 

\section{Analysis}

The algorithm above is stated in full generality, but for the sake of
analysis, we will make a couple of simplifications.  Firstly, it is
advantageous to replace the two dimensional lattice used by Kleinberg
with a one dimensional ring of vertices, and move to the directed case
where edges follow a single orientation. This means that the lattice
distance is the number of steps following the orientation of the ring
from one vertex to another. Bariere et al.  \cite{barriere:efficient}
have performed a thorough investigation of this setting. The case
$\alpha = -1$ here corresponds to the single critical, navigable case
of Kleinberg's model where greedy routing performs in $O(\log^2 n)$
steps, other values of $\alpha$ do not allow for decentralized routing
in a polylogarithmic number of steps.

Secondly we will study only the case where each vertex has exactly one
shortcut. Graphs with multiple shortcuts can be derived from this by
coalescing multiple vertices, or by slight variations in the analysis.
A final simplification of the model we analyze, shortcut independence,
will be introduced below.

\subsection{Notation}

We will index the set of vertices $V$ such that the edges of the base
graph are negatively oriented, in the sense that there is an edge from
$x$ to $x-1 \mod n$ for all $x = 0 \hdots n-1$. The function $d(x,z)$
gives the distance in this digraph from $x$ to $z$. It is not
symmetric, for example $d(x, x-1) = 1$ while $d(x-1,x) = n-1$. The
probability space used will be $V \times V \times \{E: V \mapsto V\}$
with elements $(y, z, E)$ denoting a starting point, destination, and
shortcut configuration respectively. On this we define a probability
measure $\PR$, which chooses the three elements independently, the first
two uniformly, and the third with probability defined below.

We will denote by $\ell(x, z)$ the marginal probability that $x$ has a
link to $z$. We let $D_z$ be the event that $z$ is chosen as the
destination of a query, and $H_x$ be the event that a query passes
through $x$. The conditional hitting probability of $x$ is denoted by
$h(x, z) = \PR(H_x \given D_z)$: that is $h(x,z)$ is the probability
that a query from a uniformly selected starting point with destination
$z$ passes through the point $x$. By translation invariance, both
$\ell(x, z)$ and $h(x,z)$ are functions of $d(x,z)$, and we will
sometimes see them as such (i.e. we let $\ell(x) = \ell(x,0)$ so that
$\ell(d(x,z)) = \ell(x, z)$ and define $h(x)$ equivalently.)

For sets $A \subset V$ we let $\ell(A) = \sum_{x \in A} \ell(x)$ and
$h(A) = \sum_{x \in A} \ell(x)$. We let $\tau = h(V) = \sum_{\xi =
  1}^{n-1} h(\xi)$ and note that $\tau$ is exactly the expected greedy
routing time of a query from a uniformly chosen point to 0.

\subsection{Markov Chain}

Each application of Algorithm \ref{alg:main} defines the transition of
a Markov chain on the set of shortcut configurations. Thus for any
$n$, the Markov chain in question is defined on a finite (if large)
state space. Since it can easily be seen that this chain is
irreducible and aperiodic, the chain converges to a unique stationary
distribution. The goal is to motivate that this distribution leads to
short greedy walks. The shortcut from a vertex $x$ at any time is
simply a sample of the destination of the previous walks that $x$ has
seen.  Under the stationary distribution this should not change with
time, so marginally it holds that
\begin{equation*}
\ell(x, z) = \PR(D_z \given H_x).
\end{equation*}
By using Bayes' theorem, and the definition above, we can thus write
the shortcut distribution in terms of the hitting probability:
\begin{eqnarray*}
\ell(x, z) & = & { h(x, z) \PR(D_z) \over \sum_{\xi
    \neq x} h(x, \xi)\PR(D_{\xi}) }.
\end{eqnarray*}
Since the destination is chosen uniformly at random,
$\PR(D_{\bullet})$ cancels out in numerator and denominator, and we
are left with:
\begin{equation}
\ell(x, z) = {h(x,z) \over \sum_{\xi \neq x}h(x, \xi)} = {h(x,z) \over
\sum_{\xi = 1}^{N-1}h(\xi)}
\label{eq:balanced}
\end{equation}
where the last equality follows by using translation independence and
re-indexing. In other words, $\ell(x) \propto h(x)$ for all $x$: we
will call shortcut distributions which have this property
\emph{balanced}.

\subsection{The Independent Case}

In order to get a bound on the expected greedy routing time, we will
need to make one further simplification. Instead of studying the true
stationary distribution of the rewiring process, we will look at
graphs where links are chosen independently in such a way that
(\ref{eq:balanced}) holds. There is no reason to believe that there is
independence under the true distribution (in fact, it is quite clear
that there isn't), but below we will argue heuristically that these
results are still valid.

\begin{theorem}
  For all $n \geq 1$, there exists a distribution $\ell(x)$ on $x \in
  [n-1]$ which is balanced when shortcuts are chosen independently at
  each node.
\end{theorem}

\begin{proof}
  If we consider each shortcut as chosen independently, we may view
  the query, which approaches the destination in each step, as being a
  Markov chain, and using the backwards equations for the hitting
  probability of Markov chains, we may deduce that (fixing the
  destination as 0):
\begin{eqnarray}
  h(x) = \sum_{\xi = x+1}^{N-1} h(\xi) \ell(\xi - x) + h(x+1) \sum_{\xi =
  x+2}^{N-1} \ell(\xi) + {1 \over n-1}.
\label{eq:hsys}
\end{eqnarray}
The first term above gives the probability that we enter $x$ using a
shortcut from a vertex that is $\xi$ steps from the destination, while
the second term gives the probability that we enter $x$ from the
vertex which is $x+1$ steps from the destination using the base graph.

Fix a distribution $\ell'(x)$. The hitting probability under this
distribution $h'(x)$ can be derived from (\ref{eq:hsys}), and from
this we may derive a new distribution
\[
\ell''(x) = {h'(x) \over \sum_{i=1}^{n-1}{h'(x)}}.
\]
The mapping of $\ell' \mapsto \ell''$ is continuous, since
$\sum_{i=1}^{n-1}{h'(x)} > 1$, and maps a convex set (the simplex of
$n-1$ valued distributions) into itself. By Brouwer's fix-point
theorem, there exists at least one fixpoint $\ell^*$, which by
construction is a balanced distribution.
\end{proof}

We also note that:
\begin{lemma}
  If the shortcut configuration is chosen according to a translation
  invariant distribution, then $h(x)$ is non-increasing in $x$.
\label{le:nih}
\end{lemma}

\begin{proof}
  This can been seen easily by considering any realization of the
  graph, together with a starting point, which causes a query for $0$
  to pass through $x+1$. For each such case, there is a corresponding
  configuration and starting point attained by translating each down
  one step (modulo $n$), for which a query for $0$ will pass through
  $x$.
\end{proof}

Using this we may state and prove the main result:
\begin{theorem}
\label{th:grbnd}
For every $N = 2^k$ with $k \geq 4$, let $\tau$ be the expected greedy
routing time. If shortcuts are selected independently according to a
balanced distribution at each node, then
\begin{equation*}
\tau \leq 3 k^2.
\end{equation*}
\end{theorem}

\begin{proof}
  We fix the destination as 0, and consider routing from a randomly
  chosen point. Start by dividing $1,\hdots,n-1$ into $k$ contiguous
  parts $F_1$,$F_2,\hdots,F_k$ such that
\begin{eqnarray*}
 h(F_1) \approx h(F_2) \approx \hdots \approx h(F_k)
\end{eqnarray*}
in the sense that $|h(F_i) - h(F_j)| < 2$ for all $i, j$ (such a
partition is possible since $h(x) \leq 1$ for all $x$). It follows by
proportionality that
\[
\ell(F_i) \geq {1 \over k} - {1 \over \tau} = {\tau - k \over \tau k}
\]
Let $r_0 = 0$, and $F_i = \{r_{i-1}+1, r_i\}$. We now consider a query
starting at $r_k = n-1$, the furthest point from 0 in $F_k$, and want
to find the probability that $r_k$ has a shortcut to a vertex in
$\{0,\hdots,r_{k-1}\}$. 

Assume that $r_{k-1} > r_k/2$. Then $F_k \cap \{r_k - F_k\} =
\emptyset$, so the desired probability is at least $\ell(r_k, r_k -
F_k) = \ell(F_k)$. It follows from Lemma \ref{le:nih} that the
probability of finding such a link cannot be less for any other point
in $F_k$. The expected number of steps spent in $F_k$ is thus bounded
from above by the expectation of a geometric random variable with
success probability ${\tau - k \over \tau k}$, whence
\begin{eqnarray}
  h(F_k) \leq {\tau k \over \tau - k}.
\label{eq:phbnd}
\end{eqnarray}
However, the expected time spent in each $F_i$ differs at most by a
constant, so we can conclude that:
\begin{eqnarray*}
  \tau = \sum_{i = 1}^k h(F_i) \leq 2 {\tau k^2 \over \tau - k}.
\end{eqnarray*}
which implies $\tau \leq 2 k^2 + k \leq 3 k^2$.

This leaves the case when $r_{k-1} \leq n/2$. If this holds, then by
the same reasoning, starting instead at $r_{k-1}$, we may exclude any
case but $r_{k-2} \leq r_{k-1} / 2 \leq n/4$. Continuing in this
fashion, we can exclude every case but
\[
r_1 \leq {n \over 2^{k-1}} = 2.
\]
The result then follows again since $h(F_1) \leq r_1$ and $2 < k$.
\end{proof}

\subsection{Dependencies}

In order to fully prove that the rewiring algorithm presented above
leads to a navigable graph, one needs to prove that the dependencies
present in the resulting distribution of shortcuts are not destructive
to the argument. In fact, our reasoning uses independence only at one
point. In the proof of Theorem \ref{th:grbnd}, after having calculated
a marginal bound of $\approx 1/k$ of the probability that each point
in $F_i$ has a shortcut to a point outside the phase but closer to 0,
we conclude (\ref{eq:phbnd}): that this means that the expected number
of steps in $F_i$ is at most $\approx k$. This is true only if we draw
a shortcut independently at each vertex in $F_i$ that we reach, or if
conditioned on not having found a useful shortcut in one step, the
probability of doing so in the next increases.

Proving the full result formally is still an open problem, but one can
see heuristically why it should hold. There are two forms of
dependence present between edges created using the destination
sampling algorithm. The first comes from the fact that two edges may
have been created from the same query, and thus have the same
destination. The parameter $p$ is introduced into the algorithm to
alleviate this (if $p$ is large, one can very clearly see that nearby
vertices tend to have the same shortcut destination, with considerable
cost to routing performance), and by choosing $p$ sufficiently small,
we can make it negligible. The second type of dependence comes from
the fact that what other edges are present around a vertex $x$ will,
of course, greatly affect the probability of whether a query for some
vertex $z$ passes through $x$.

When trying to bound the expected time spent in each $F_i$ in the
proof of Theorem \ref{th:grbnd}, and thus calculating the probability
that $x$ has a shortcut that takes the query out of $F_i$, we have to
condition on the previously encountered vertices having shortcuts that
failed to do this. These could either be shortcuts from one earlier
vertex in $F_i$ to another, or shortcuts that overshoot the target (0)
and thus are not used by the query. The presence of neither type of
shortcut would seem to make it less likely that a query for a point
in $A = \{0, \hdots, r_{i-1}\}$ passes through $x$, and hence one
would not expect that the conditioning should make $h(x,A)$ (and
thus $\ell(x, A)$) smaller.  Formalizing this argument has, however,
proved difficult.

\section{Simulation}

Simulations indicate that the algorithm gives results which scale as
desired in the number of greedy steps, and that the resulting shortcut
distribution approximates $1 / \log(n) d(x,y)$ as expected.

The results in the directed one-dimensional case can be seen in Figure
\ref{fig:grdist}. To get these results, the graph is started with no
shortcuts, and then the algorithm is run $10 N$ times to initialize
the references. The value of $p = 0.1$ is used. The greedy distance is
then measured as the average of 100,000 walks, each updating the graph
according to the algorithm (this decreases the variance of the
estimate).

The square root of the mean greedy distance increases linearly as the
graph size increases exponentially, just as we would expect. In fact
our algorithm leads to better simulation results than choosing from
Kleinberg's distribution. Doubling the graph size is found to increase
the square route of the greedy distance by $\approx 0.41$ when links are
selected using our algorithm, compared to an increase of $\approx 0.51$
when Kleinberg's model is used. (For Kleinberg's model we can use
(\ref{eq:hsys}) to calculate numerically exact values for $\tau$,
allowing us to confirm this figure.)

\begin{figure}
\begin{center}
\input{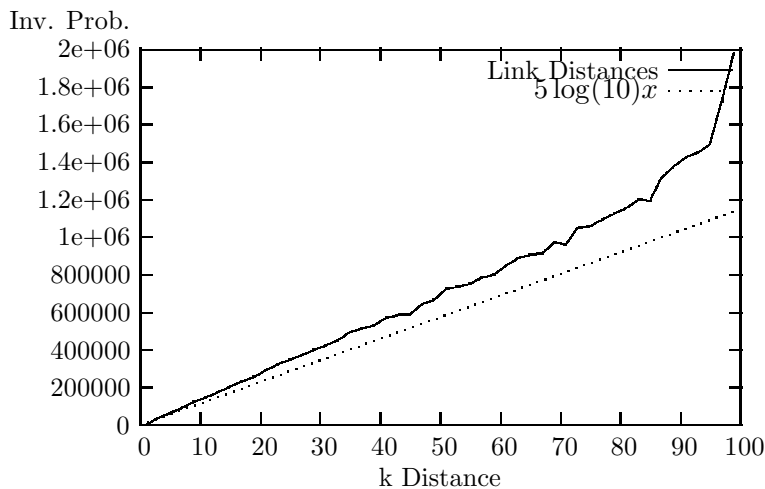}
\caption{The inverse of distribution of shortcut distances, with $N = 100000$, $p =
  0.10$. The straight line is the inverse of the harmonic distribution.}
\label{fig:linkdist}
\end{center}
\end{figure}

In Figure \ref{fig:linkdist} the marginal distribution of shortcut
lengths is plotted. It is roughly harmonic in shape, except that it
creates less links of length close to the size of the graph. This
may be part of the reason why it is able to outperform Kleinberg's
model: while Kleinberg's model is asymptotically correct, this
algorithm takes into account finite size effects. (This reasoning is
similar to that of the authors of \cite{clauset:navig}. Like them, we
have no strong analytic arguments for why this should be the case,
which makes it a tenuous argument at best.)

\begin{figure}
\begin{center}
\setlength{\unitlength}{0.180675pt}

\ifx\plotpoint\undefined\newsavebox{\plotpoint}\fi
\begin{picture}(1500,900)(0,0)
\sbox{\plotpoint}{\rule[-0.200pt]{0.400pt}{0.400pt}}%
\put(141.0,123.0){\rule[-0.200pt]{4.818pt}{0.400pt}}
\put(121,123){\makebox(0,0)[r]{ 3}}
\put(1419.0,123.0){\rule[-0.200pt]{4.818pt}{0.400pt}}
\put(141.0,246.0){\rule[-0.200pt]{4.818pt}{0.400pt}}
\put(121,246){\makebox(0,0)[r]{ 4}}
\put(1419.0,246.0){\rule[-0.200pt]{4.818pt}{0.400pt}}
\put(141.0,369.0){\rule[-0.200pt]{4.818pt}{0.400pt}}
\put(121,369){\makebox(0,0)[r]{ 5}}
\put(1419.0,369.0){\rule[-0.200pt]{4.818pt}{0.400pt}}
\put(141.0,491.0){\rule[-0.200pt]{4.818pt}{0.400pt}}
\put(121,491){\makebox(0,0)[r]{ 6}}
\put(1419.0,491.0){\rule[-0.200pt]{4.818pt}{0.400pt}}
\put(141.0,614.0){\rule[-0.200pt]{4.818pt}{0.400pt}}
\put(121,614){\makebox(0,0)[r]{ 7}}
\put(1419.0,614.0){\rule[-0.200pt]{4.818pt}{0.400pt}}
\put(141.0,737.0){\rule[-0.200pt]{4.818pt}{0.400pt}}
\put(121,737){\makebox(0,0)[r]{ 8}}
\put(1419.0,737.0){\rule[-0.200pt]{4.818pt}{0.400pt}}
\put(141.0,860.0){\rule[-0.200pt]{4.818pt}{0.400pt}}
\put(121,860){\makebox(0,0)[r]{ 9}}
\put(1419.0,860.0){\rule[-0.200pt]{4.818pt}{0.400pt}}
\put(141.0,123.0){\rule[-0.200pt]{0.400pt}{4.818pt}}
\put(141,82){\makebox(0,0){-2}}
\put(141.0,840.0){\rule[-0.200pt]{0.400pt}{4.818pt}}
\put(401.0,123.0){\rule[-0.200pt]{0.400pt}{4.818pt}}
\put(401,82){\makebox(0,0){ 0}}
\put(401.0,840.0){\rule[-0.200pt]{0.400pt}{4.818pt}}
\put(660.0,123.0){\rule[-0.200pt]{0.400pt}{4.818pt}}
\put(660,82){\makebox(0,0){ 2}}
\put(660.0,840.0){\rule[-0.200pt]{0.400pt}{4.818pt}}
\put(920.0,123.0){\rule[-0.200pt]{0.400pt}{4.818pt}}
\put(920,82){\makebox(0,0){ 4}}
\put(920.0,840.0){\rule[-0.200pt]{0.400pt}{4.818pt}}
\put(1179.0,123.0){\rule[-0.200pt]{0.400pt}{4.818pt}}
\put(1179,82){\makebox(0,0){ 6}}
\put(1179.0,840.0){\rule[-0.200pt]{0.400pt}{4.818pt}}
\put(1439.0,123.0){\rule[-0.200pt]{0.400pt}{4.818pt}}
\put(1439,82){\makebox(0,0){ 8}}
\put(1439.0,840.0){\rule[-0.200pt]{0.400pt}{4.818pt}}

\put(141.0,123.0){\rule[-0.200pt]{234.516pt}{0.400pt}}

\put(1439.0,123.0){\rule[-0.200pt]{0.400pt}{133.167pt}}

\put(141.0,860.0){\rule[-0.200pt]{234.516pt}{0.400pt}}

\put(141.0,123.0){\rule[-0.200pt]{0.400pt}{133.167pt}}

\put(140,920){\makebox(0,0){Sqrt. Mean Pathlength}}
\put(790,21){\makebox(0,0){$\log_2$ of $N / 10000$}}
\put(1279,820){\makebox(0,0)[r]{algorithm}}
\put(1299.0,820.0){\rule[-0.200pt]{24.090pt}{0.400pt}}
\put(141,184){\usebox{\plotpoint}}
\multiput(141.00,184.58)(1.481,0.499){173}{\rule{1.282pt}{0.120pt}}
\multiput(141.00,183.17)(257.340,88.000){2}{\rule{0.641pt}{0.400pt}}
\multiput(401.00,272.58)(1.411,0.499){181}{\rule{1.226pt}{0.120pt}}
\multiput(401.00,271.17)(256.455,92.000){2}{\rule{0.613pt}{0.400pt}}
\multiput(660.00,364.58)(1.416,0.499){181}{\rule{1.230pt}{0.120pt}}
\multiput(660.00,363.17)(257.446,92.000){2}{\rule{0.615pt}{0.400pt}}
\multiput(920.00,456.58)(1.380,0.499){185}{\rule{1.202pt}{0.120pt}}
\multiput(920.00,455.17)(256.505,94.000){2}{\rule{0.601pt}{0.400pt}}
\multiput(1179.00,550.58)(1.371,0.499){187}{\rule{1.195pt}{0.120pt}}
\multiput(1179.00,549.17)(257.520,95.000){2}{\rule{0.597pt}{0.400pt}}
\put(1279,779){\makebox(0,0)[r]{harmonic}}
\multiput(1299,779)(20.756,0.000){5}{\usebox{\plotpoint}}
\put(1399,779){\usebox{\plotpoint}}
\put(141,214){\usebox{\plotpoint}}
\multiput(141,214)(19.089,8.149){14}{\usebox{\plotpoint}}
\multiput(401,325)(18.970,8.423){14}{\usebox{\plotpoint}}
\multiput(660,440)(18.818,8.757){14}{\usebox{\plotpoint}}
\multiput(920,561)(18.721,8.963){13}{\usebox{\plotpoint}}
\multiput(1179,685)(18.678,9.052){14}{\usebox{\plotpoint}}
\put(1439,811){\usebox{\plotpoint}}
\end{picture}
\setlength{\unitlength}{0.240900pt}

\caption{The expected greedy walk length of the selection algorithm,
  compared to selection according to harmonic distances, in a two
  dimensional base grid.}
\label{fig:grdist2d}
\end{center}
\end{figure}
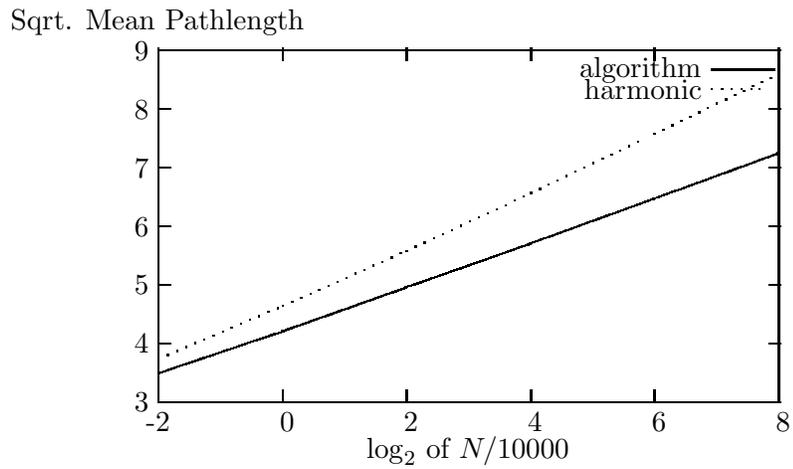

The algorithm has also been simulated to good effect using base graphs
of higher dimensions. Figure \ref{fig:grdist2d} shows the mean greedy
distance for two dimensional grids of increasing size. Here also, the
algorithm creates configurations that seem to display square
logarithmic growth, and which perform considerably better than
explicit selection according to Kleinberg's model.

\begin{figure}
\begin{center}
\setlength{\unitlength}{0.180675pt}
\ifx\plotpoint\undefined\newsavebox{\plotpoint}\fi
\begin{picture}(1500,900)(0,0)
\font\gnuplot=cmr10 at 10pt
\gnuplot
\sbox{\plotpoint}{\rule[-0.200pt]{0.400pt}{0.400pt}}%
\put(161.0,123.0){\rule[-0.200pt]{4.818pt}{0.400pt}}
\put(141,123){\makebox(0,0)[r]{ 4}}
\put(1419.0,123.0){\rule[-0.200pt]{4.818pt}{0.400pt}}
\put(161.0,246.0){\rule[-0.200pt]{4.818pt}{0.400pt}}
\put(141,246){\makebox(0,0)[r]{ 5}}
\put(1419.0,246.0){\rule[-0.200pt]{4.818pt}{0.400pt}}
\put(161.0,369.0){\rule[-0.200pt]{4.818pt}{0.400pt}}
\put(141,369){\makebox(0,0)[r]{ 6}}
\put(1419.0,369.0){\rule[-0.200pt]{4.818pt}{0.400pt}}
\put(161.0,492.0){\rule[-0.200pt]{4.818pt}{0.400pt}}
\put(141,492){\makebox(0,0)[r]{ 7}}
\put(1419.0,492.0){\rule[-0.200pt]{4.818pt}{0.400pt}}
\put(161.0,614.0){\rule[-0.200pt]{4.818pt}{0.400pt}}
\put(141,614){\makebox(0,0)[r]{ 8}}
\put(1419.0,614.0){\rule[-0.200pt]{4.818pt}{0.400pt}}
\put(161.0,737.0){\rule[-0.200pt]{4.818pt}{0.400pt}}
\put(141,737){\makebox(0,0)[r]{ 9}}
\put(1419.0,737.0){\rule[-0.200pt]{4.818pt}{0.400pt}}
\put(161.0,860.0){\rule[-0.200pt]{4.818pt}{0.400pt}}
\put(141,860){\makebox(0,0)[r]{ 10}}
\put(1419.0,860.0){\rule[-0.200pt]{4.818pt}{0.400pt}}
\put(161.0,123.0){\rule[-0.200pt]{0.400pt}{4.818pt}}
\put(161,82){\makebox(0,0){ 0}}
\put(161.0,840.0){\rule[-0.200pt]{0.400pt}{4.818pt}}
\put(417.0,123.0){\rule[-0.200pt]{0.400pt}{4.818pt}}
\put(417,82){\makebox(0,0){ 2}}
\put(417.0,840.0){\rule[-0.200pt]{0.400pt}{4.818pt}}
\put(672.0,123.0){\rule[-0.200pt]{0.400pt}{4.818pt}}
\put(672,82){\makebox(0,0){ 4}}
\put(672.0,840.0){\rule[-0.200pt]{0.400pt}{4.818pt}}
\put(928.0,123.0){\rule[-0.200pt]{0.400pt}{4.818pt}}
\put(928,82){\makebox(0,0){ 6}}
\put(928.0,840.0){\rule[-0.200pt]{0.400pt}{4.818pt}}
\put(1183.0,123.0){\rule[-0.200pt]{0.400pt}{4.818pt}}
\put(1183,82){\makebox(0,0){ 8}}
\put(1183.0,840.0){\rule[-0.200pt]{0.400pt}{4.818pt}}
\put(1439.0,123.0){\rule[-0.200pt]{0.400pt}{4.818pt}}
\put(1439,82){\makebox(0,0){ 10}}
\put(1439.0,840.0){\rule[-0.200pt]{0.400pt}{4.818pt}}

\put(161.0,123.0){\rule[-0.200pt]{230.25pt}{0.400pt}}

\put(1439.0,123.0){\rule[-0.200pt]{0.400pt}{133.167pt}}

\put(161.0,860.0){\rule[-0.200pt]{230.25pt}{0.400pt}}

\put(161.0,123.0){\rule[-0.200pt]{0.400pt}{133.167pt}}

\put(140,920){\makebox(0,0){Sqrt. Mean Pathlength}}
\put(800,21){\makebox(0,0){$\log_2$ of $N / 1000$}}

\put(1279,820){\makebox(0,0)[r]{algorithm}}
\put(1299.0,820.0){\rule[-0.200pt]{24.090pt}{0.400pt}}
\put(161,172){\usebox{\plotpoint}}
\multiput(161.00,172.58)(1.212,0.498){103}{\rule{1.066pt}{0.120pt}}
\multiput(161.00,171.17)(125.787,53.000){2}{\rule{0.533pt}{0.400pt}}
\multiput(289.00,225.58)(1.212,0.498){103}{\rule{1.066pt}{0.120pt}}
\multiput(289.00,224.17)(125.787,53.000){2}{\rule{0.533pt}{0.400pt}}
\multiput(417.00,278.58)(1.202,0.498){103}{\rule{1.058pt}{0.120pt}}
\multiput(417.00,277.17)(124.803,53.000){2}{\rule{0.529pt}{0.400pt}}
\multiput(544.00,331.58)(1.235,0.498){101}{\rule{1.085pt}{0.120pt}}
\multiput(544.00,330.17)(125.749,52.000){2}{\rule{0.542pt}{0.400pt}}
\multiput(672.00,383.58)(1.259,0.498){99}{\rule{1.104pt}{0.120pt}}
\multiput(672.00,382.17)(125.709,51.000){2}{\rule{0.552pt}{0.400pt}}
\multiput(800.00,434.58)(1.259,0.498){99}{\rule{1.104pt}{0.120pt}}
\multiput(800.00,433.17)(125.709,51.000){2}{\rule{0.552pt}{0.400pt}}
\multiput(928.00,485.58)(1.259,0.498){99}{\rule{1.104pt}{0.120pt}}
\multiput(928.00,484.17)(125.709,51.000){2}{\rule{0.552pt}{0.400pt}}
\multiput(1056.00,536.58)(1.249,0.498){99}{\rule{1.096pt}{0.120pt}}
\multiput(1056.00,535.17)(124.725,51.000){2}{\rule{0.548pt}{0.400pt}}
\multiput(1183.00,587.58)(1.285,0.498){97}{\rule{1.124pt}{0.120pt}}
\multiput(1183.00,586.17)(125.667,50.000){2}{\rule{0.562pt}{0.400pt}}
\multiput(1311.00,637.58)(1.285,0.498){97}{\rule{1.124pt}{0.120pt}}
\multiput(1311.00,636.17)(125.667,50.000){2}{\rule{0.562pt}{0.400pt}}
\put(1279,779){\makebox(0,0)[r]{harmonic}}
\multiput(1299,779)(20.756,0.000){5}{\usebox{\plotpoint}}
\put(1399,779){\usebox{\plotpoint}}
\put(161,222){\usebox{\plotpoint}}
\multiput(161,222)(19.384,7.420){7}{\usebox{\plotpoint}}
\multiput(289,271)(18.737,8.929){7}{\usebox{\plotpoint}}
\multiput(417,332)(19.154,7.994){7}{\usebox{\plotpoint}}
\multiput(544,385)(17.969,10.388){7}{\usebox{\plotpoint}}
\multiput(672,459)(18.506,9.398){7}{\usebox{\plotpoint}}
\multiput(800,524)(19.015,8.319){6}{\usebox{\plotpoint}}
\multiput(928,580)(18.506,9.398){7}{\usebox{\plotpoint}}
\multiput(1056,645)(18.476,9.456){7}{\usebox{\plotpoint}}
\multiput(1183,710)(18.793,8.809){7}{\usebox{\plotpoint}}
\multiput(1311,770)(18.737,8.929){7}{\usebox{\plotpoint}}
\put(1439,831){\usebox{\plotpoint}}
\end{picture}
\setlength{\unitlength}{0.240900pt}
\caption{The expected
  greedy walk length using our selection algorithm, compared to
  selection according the harmonic distribution, in a directed ring.}
\label{fig:grdist}
\end{center}
\end{figure}
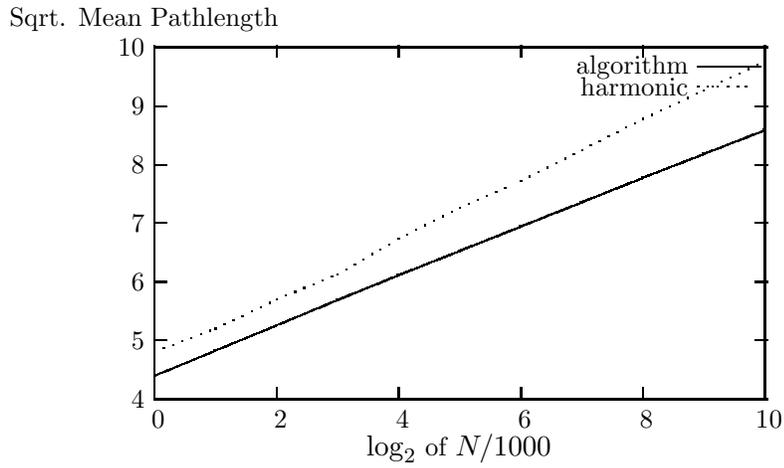

\section{Conclusion}

We have introduced an evolutionary model that by successively updating
the the shortcut edges of a small-world graph creates configurations
which are navigable. We have explored this model both
analytically and with the help of simulation, and found support for
the claim that navigability should arise.

The major open question is to complete the rigorous analysis of the
stationary distribution, in particularly with regard to the
dependencies between shortcuts. Random graphs with dependencies
between the edges are notoriously difficult to analyze mathematically,
and possibility of doing so usually relies on finding a formulation
where the edges can be seen to be independent conditioned on a certain
event (already Kleinberg's model is an example of this: the edges do
not exist independently, but are independent conditioned on the
position of the nodes).  No such formulation has been found here, but
the existence cannot be ruled out.

Further, there are interesting questions regarding the scope of the
results. We have a upper bound for the navigable case which matches
Kleinberg, but it would be interesting to see if lower bounds can be
found for the case when $\ell(x, z)$ deviates from greatly from
proportionality to $h(x,z)$, in the notation above. While it seems
clear that this must be the case, a direct proof would be
illustrative. Finally, it is noted that the destination sampling
algorithm suggested can be stated and implemented independently of the
structure of the underlying graph (and thus distance function), and
there is no reason to believe it wouldn't work with just about any
graph.  Exploring the limits of the algorithms applicability is an
interesting, open problem.


\newpage
\bibliographystyle{plain}
\bibliography{../tex/ossa}

\end{document}